\documentclass{an}
\usepackage{graphicx}
\usepackage{times}
\usepackage[british]{babel}
\begin{document}
% The following seven commands are intended for editorial usage and should be ignored by
% the author(s).
\Pagespan{1}{}% Document's page range. 
% If second parameter is left empty, the last page is computed automatically.
\Yearpublication{}%
\Yearsubmission{}%
\Month{}%   
\Volume{}%  
\Issue{}% 
% \DOI{This.is/not.aDOI}% 

\title{The Blazhko behavior of RV UMa}

\author{Zs., Hurta\inst{1,2}
}
\titlerunning{The Blazhko behavior of RV UMa}
\authorrunning{Zs., Hurta}
\institute{
E\"otv\"os Lor\'and University, Department of Astronomy, P.O. Box 32, H-1518 Budapest, Hungary
\and
Konkoly Observatory of the Hungarian Academy of Sciences, P.O. Box 67, H-1525 Budapest, Hungary
}

\received{30 May 2005}
\accepted{11 Nov 2005}
\publonline{later}


\abstract{%
RV UMa is one of the most extensively studied RR Lyrae stars showing Blazhko modulation. Its photometric observations cover more than 90 years. The published photoelectric observations of RV UMa obtained at the Konkoly \hbox{Observatory} (Kany\'o, 1976) were re-considered and completed with previously unpublished data. During the time interval of the ob\-ser\-va\-tions the periods of both the pulsation and the modulation varied within the ranges of 0.000007 and 0.9 days, respectively. We have found a definite but not strict inverse relation between the pulsation and modulation periods of RV~UMa.
}

\maketitle
\sloppy

\section{Introduction}
RV UMa is one of the fundamental mode RR Lyrae stars showing large amplitude light curve modulation (Blazhko effect, see S\'odor 2007, in this issue), which has been regularly observed during the last century. The available comprehensive photometric data of RV UMa make it possible to follow long term changes in its pulsation and modulation properties, that may help to solve the hundred-year old puzzle of the modulation phenomenon.

Bal\'azs \& Detre (1957) studied the changes of the pulsation and modulation periods of the star most thoroughly. They found that the changes in the periods were the opposite of each other. This behavior was later confirmed by Kany\'o (1976). A detailed, quantified analysis of the long term behavior of RV UMa has not, however, been performed yet.

In this paper, based on all the available photometric observations, we show how the pulsation and modulation periods of RV UMa changed during the 100-year time base.

\section{The Data}
\begin{table*}[t]
\begin{tabular}{cccrrrrrrc}
\hline
ID & Type & JD & Night & Data & \multicolumn{2}{c}{Light curve} & \multicolumn{2}{c}{b max} & References\\
 & & [JD$-$2400000] & & & \multicolumn{1}{c}{$a$} & \multicolumn{1}{c}{$b$} & \multicolumn{1}{c}{$a$} & \multicolumn{1}{c}{$b$} & \\
\hline
$1$ & $vis$ & $17852-18764$ & $47$ & $323$ & $0.13$ & $9.43$ & $0.22$ & $8.96$ & Ishchenko 1939 \\
$2$ & $vis$ & $17958-18222$ & $41$ & $104$ & $-0.13$ & $12.35$ & $-0.23$ & $14.07$ & Luizet 1909 \\
$3$ & $vis$ & $19757-20527$ & $171$ & $236$ & $1.39$ & $-3.72$ & $1.05$ & $-0.32$ & Nijland 1922 \\
$4$ & $vis$ & $23307-23311$ & $4$ & $27$ & $0.13$ & $9.43$ & $0.22$ & $8.96$ & Ishchenko 1939 \\
$5$ & $vis$ & $29044-29071$ & $8$ & $200$ & $0.90$ & $1.54$ & $0.90$ & $1.54$ & Ishchenko 1939 \\
$6$ & $vis$ & $32998-34273$ & $88$ & $485$ & $1.48$ & $-4.94$ & $0.72$ & $2.89$ & Yudkina 1951, 1953\\
$7$ & $pg$ & $20238-20493$ & $5$ & $70$ & $1.06$ & $0.05$ & $1.06$ & $0.05$ & Beljawsky 1915\\
$8$ & $pg$ & $20482-20985$ & $16$ & $281$ & $0.95$ & $1.03$ & $0.97$ & $0.62$ & Jordan 1920\\
$9$ & $pg$ & $23520-24373$ & $24$ & $167$ & $0.91$ & $1.69$ & $0.74$ & $3.27$ & Subbotin 1927\\
$10$ & $pg$ & $25066-25441$ & $28$ & $314$ & $1.03$ & $0.62$ & $0.76$ & $3.07$ & Scharonow 1930\\
$11$ & $pg$ & $28322-28691$ & $12$ & $376$ & $1.04$ & $-0.14$ & $0.99$ & $0.39$ & Bal\'azs \& Detre 1957 \\
$12$ & $pg$ & $31911-34195$ & $18$ & $630$ & $1.04$ & $-0.14$ & $0.99$ & $0.39$ & Bal\'azs \& Detre 1957 \\
$13$ & $pe_{\rm{unf}}$ & $35566-36020$ & $35$ & $756$ & $1.14$ & $-1.31$ & $1.02$ & $-0.09$ & Bal\'azs \& Detre 1957 \\
$14$ & $pe_V$ & $36647-37061$ & $18$ & $448$ & $1.32$ & $-3.11$ & $1.32$ & $-3.11$ & Preston \& Spinrad 1967 \\
$15$ & $pe_B$ & $36647-37061$ & $18$ & $448$ & $1.00$ & $0.00$ & $1.00$ & $0.00$ & Preston \& Spinrad 1967 \\
$16$ & $pe_V$ & $36229-42422$ & $84$ & $3077$ & $1.32$ & $-3.11$ & $1.32$ & $-3.11$ & Konkoly observations$^*$ \\
$17$ & $pe_B$ & $36229-42422$ & $91$ & $3325$ & $1.00$ & $0.00$ & $1.00$ & $0.00$ & Konkoly observations$^*$ \\
$18$ & $CCD$ & $47871-49018$ & $41$ & $128$ & $1.21$ & $-2.05$ & $1.24$ & $-2.33$ & Hipparcos 1997 \\
$19$ & $CCD$ & $51274-51633$ & $107$ & $423$ & $1.48$ & $-5.25$ & $1.34$ & $-3.83$ & Wo\'zniak et al. 2004 \\
\hline
\end{tabular}\\
$^*$These data were partly published by Kany\'o (1976), a corrected data set is used in this paper (see details in the text).
\caption{Photometric observation of RV UMa, and the coefficients of the linear transformations of the light curves and maximum brightness data which were applied in order to yield a complete, homogeneous data set.}
\label{tab1}
\end{table*}
All the published data of RV UMa which were suitable for the analysis of the light curve and/or maximum brightness variation were collected and used. The summary of the collected photometries is given in Table~\ref{tab1}. We did not use the published maximum brightness times and magnitudes of these data sets, instead we have derived individual maximum times and brightness values from the original data in a homogeneous manner. If the data distribution did not allow to define accurate enough individual maximum timings and brightness values then normal maxima were determined from points of about 10 day intervals, during that interval the amplitude of the pulsation does not change significantly taking into account the 90 day long periodicity of the modulation. The derived timings and transformed magnitudes (see details later) of the light maxima and timings of maximum amplitude epochs of the modulation are available electronically at http://konkoly.hu/24/publications/rvuma/.

Constructing the $O-C$ diagram of the pulsation we also utilized maximum times from the literature when no photometric data were given (Agerer, Dahm \& H\"ubscher 1999, 2001; Agerer \& H\"ubscher 2002, 2003; Ahnert 1961; Aubaud 1991; Braune \& H\"ubscher 1967, 1987; Braune \& Mundry 1982; Braune, H\"ubscher \& Mundry 1970, 1972, 1977, 1979; Dombrovski 1935; Fitch, Wisniewski \& Johnson 1966; Geyer 1961; H\"ubscher 2000, 2001, 2003, 2005; H\"ubscher et al. 1994, 1998, 1999; H\"ubscher \& Lichtenknecker 1988; H\"ubscher \& Mundry 1984; H\"ubscher, Agerer \& Wunder 1991, 1992, 1995; H\"ubscher, Lichtenknecker \& Meyer 1987; H\"ubscher, Lichtenknecker \& Wunder 1989, 1990; H\"ubscher, Paschke \& Walter 2005; Lange \& Kanishcheva 1961; Le Borgne, Klotz \& Bo\"er 2004, 2005; Solo-viev 1936, 1941; Tsesevich 1969; Vandenbroere 1997,1999, 2001, 2005). 

The photoelectric observations obtained at the Konkoly Observatory and partly published by Kany\'o (1976) were re-considered. We have found that the published data were erroneously transformed to the standard system which explains the magnitude differences between the Preston \& Spinrad (1967) and Kany\'o (1976) magnitudes. We have rereduced the original observations to correct this error. In the new reduction we dropped out the observations of two nights (JD 2\,437\,672, JD 2\,437\,780) and some uncertain data points of other nights. Altogether 43 data points were left out from the Kany\'o's (1976) data. We completed these data with further, previously unpublished observations of our Observatory which extended the observational material with data of 33 nights, 835 data points in the $V$ band and 40 nights, 1083 data points in the $B$ band. The expanded Konkoly observations cover nearly 17 years which gives us the opportunity to determine more accurate pulsation and modulation periods for a longer time interval.

\begin{figure}
\begin{center}
\includegraphics[width=5.3cm,angle=-90]{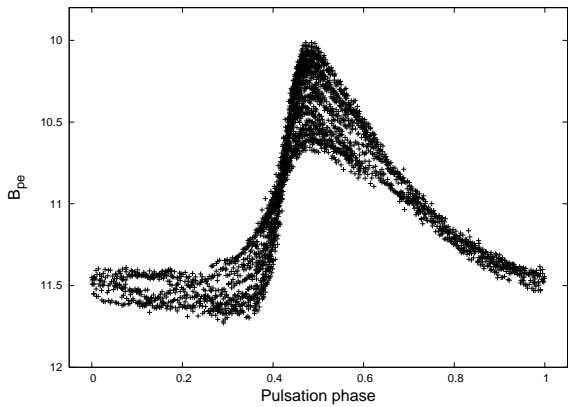}
\end{center}
\caption{Folded light curves of the photoelectric $B$ observations obtained at the Konkoly Observatory between JD 2\,436\,229 and 2\,442\,422. Data are plotted according to the phases of the pulsation. The effect of the modulation concentrates on the minima and the maxima of the light curve, while on the rising and ascending branches `fix points' exist. When the minima is the brightest then the maxima is the faintest and when the minima is the faintest then the maxima is the brightest.}
\label{fig1}
\end{figure}

\begin{figure}
\begin{center}
\includegraphics[width=5.3cm,angle=-90]{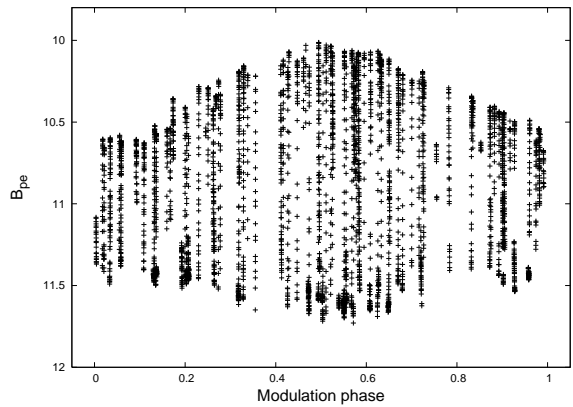}
\end{center}
\caption{Folded light curves of the photoelectric $B$ observations obtained at the Konkoly Observatory. Data are plotted according to the phases of the Blazhko modulation. The data almost completely cover each phase of the modulation. The amplitude of the changes in the minima is smaller than the changes of the maxima. The scatter of the maximum brightness is the result of the changes in the period of the modulation during the observations (see also Figure~\ref{fig45}).}
\label{fig2}
\end{figure}

Figures~\ref{fig1} and~\ref{fig2} show the photoelectric $B$ observations obtained at our Observatory phased with $0\fd468062$ pulsation and $90\fd44$ modulation periods, respectively.


In order to extract the most information possible from the observations, all the photometric data were transformed to the magnitude scale of the photoelectric $B$ observations. Although no exact linear transformation exists between the light curves in different wavelength bands, the linearly transformed, `homogenized' data have a much better time coverage than the original data sets, allowing to determine the pulsation and modulation periods at different epochs more accurately. The different observations were homogenized using linear transformations derived separately for the light curve data and for the maximum brightness magnitudes. The light curve data were transformed by the fit to the mean value of the brightness of minima and maxima to those of the standard light curve, while the maximum brightness magnitudes were transformed by the fit to the minima and maxima of the standard maximum brightness magnitudes. As standard light curve and standard maximum brightness magnitudes, we used the photoelectric $B$ observations obtained at the Konkoly Observatory, because they were the most numerous and they almost completely covered each phase of the Blazhko cycle (Figure~\ref{fig2}). In deriving the linear coefficients of the transformations it was also taken into account which Blazhko phases were or were not covered by the data. The coefficients of the linear transformations applied are listed in Table~\ref{tab1}. The linearly transformed maximum brightness data are also given in the electronic tables of the maximum timings. The differences between the coefficients ``a" of the light curve and of the brightness of maxima are relatively small for most of the data sets which justify the reliability of our method. In the case of visual observations the discrepancy can be explained by their considerably larger uncertainties. The ``b" coefficients stand for the zero points' differences. This explains their large divergences.

\begin{figure}
\begin{center}
\includegraphics[width=5.1cm,angle=-90]{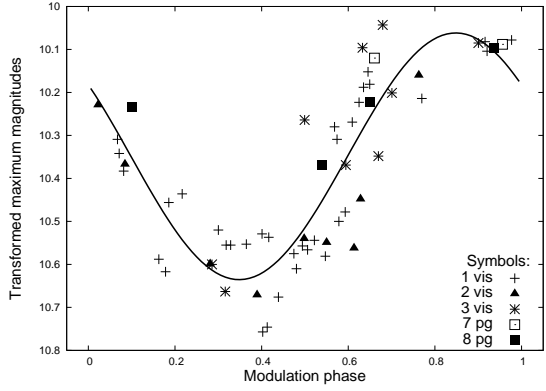}
\end{center}
\caption{The transformed magnitudes of maxima  between JD 2\,417\,852 and 2\,420\,985 folded with the best Blazhko period. Different observations are shown using different symbols. The numbers refer to Table~\ref{tab1}.}
\label{fig3}
\end{figure}

As an example Figure~\ref{fig3} shows the maximum brightness variation of the homogenized data set for the JD 2\,417\,852 $-$ 2\,420\,985 interval. Observations on different magnitude scales well define the same maximum brightness variation during the Blazhko cycles after the linear transformation. 

\section{Data analysis}

\begin{figure}
\begin{center}
\includegraphics[angle=-90,width=0.95\hsize]{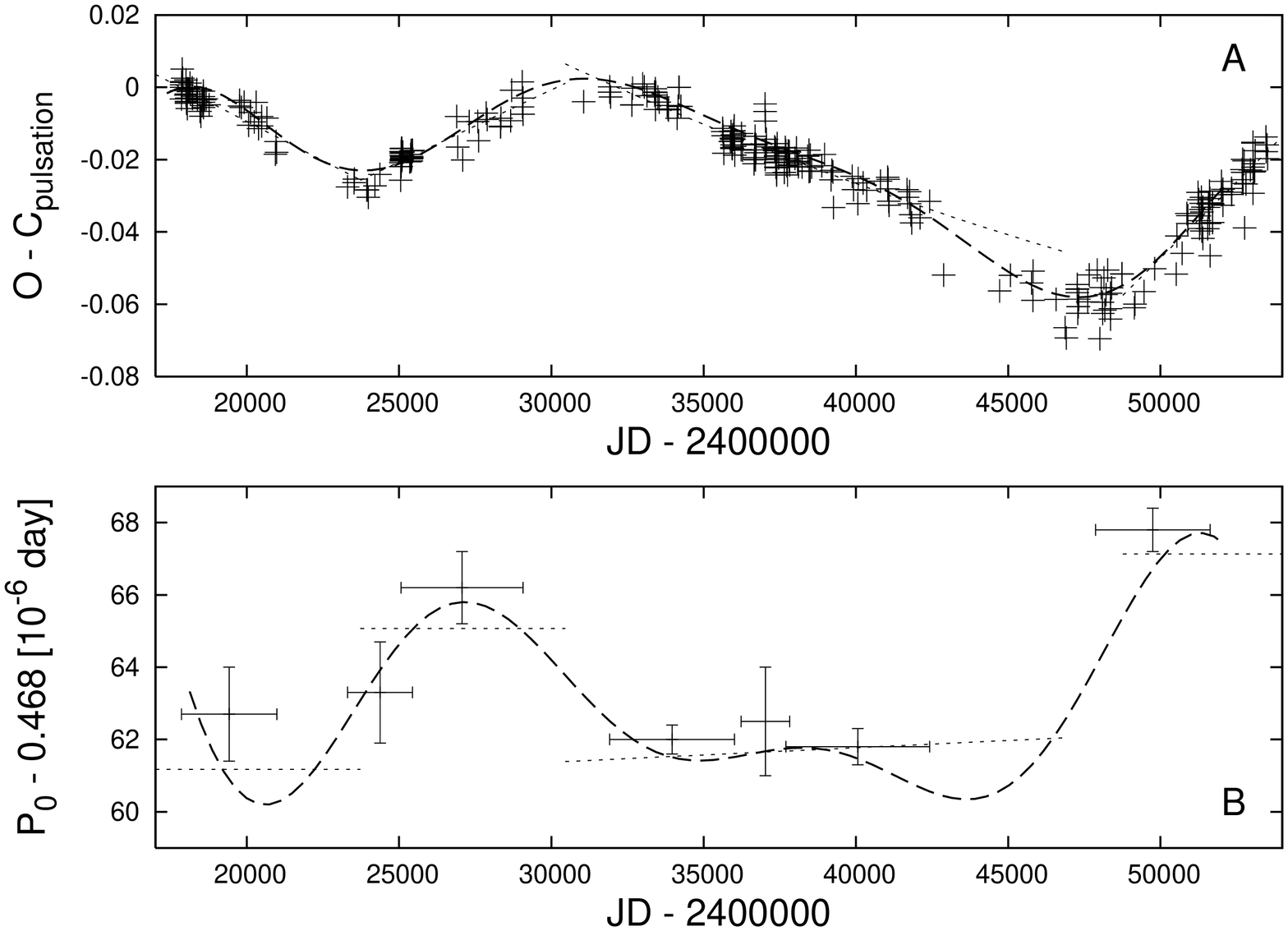}
\includegraphics[angle=-90,width=0.95\hsize]{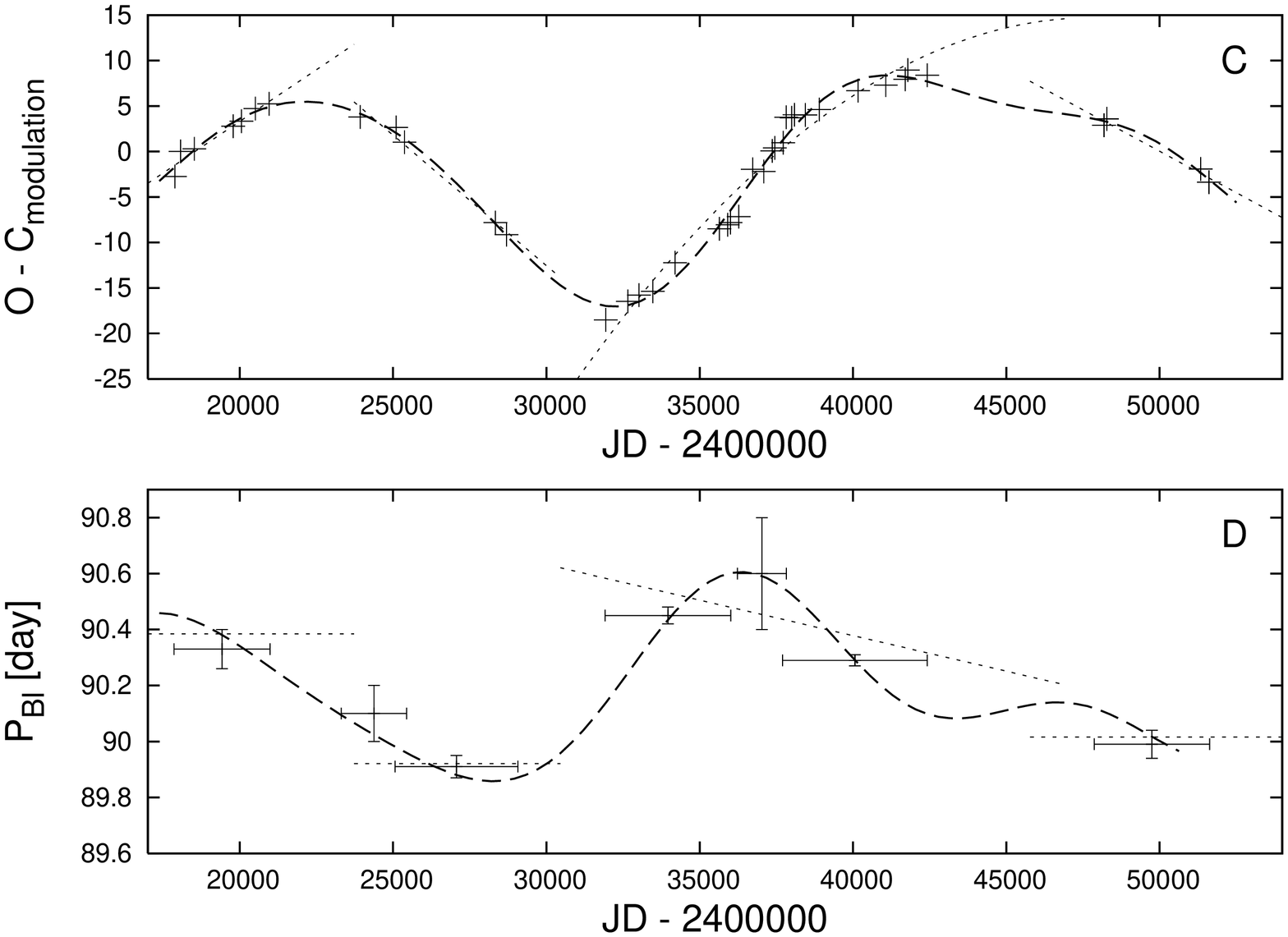}
\end{center}
\caption{Panels A and B show the pulsation $O-C$ diagram of maximum brightness and the directly measured pulsation period values listed in Table~\ref{tab2}. Panels C and D show the plots of the modulation $O-C$ diagram of the maximum amplitudes and the measured Blazhko periods. The fits in panels B and D are not the actual fits to the data but correspond to the derivatives of the harmonic (dashed lines) and low order polynomial (dotted lines) fits to the $O-C$ data shown in panels A and C.}
\label{fig45}
\end{figure}

Within the uncertainties arising from the differences of the photometries we did not find any sign of changes in the pulsation and modulation properties of RV UMa except the detectable changes in the modulation period. Therefore our investigation is focused on the connected changes in the pulsation and modulation periods on the 100-year time base.

Period changes were followed by the help of the $O-C$ plots of the pulsation light and modulation amplitude maxima (A and C panels in Figure~\ref{fig45}). The $O-C$ data were calculated according to the following ephemerides:

\noindent
$t_\mathrm{pulsation\ max} = J.D. 2417861.907 + 0.468063203 \times E_{\rm{p}}$

\noindent
$t_\mathrm{modulation\ max} = J.D. 2418065.8 + 90.18 \times E_{\rm{Bl}}$

The $O-C$ data were fitted by different order polynomial and harmonic functions and the period changes were defined as the derivatives of the different fits to the $O-C$ data. In Figure~\ref{fig45} the dashed and dotted lines correspond to first or second order polynomial fits to the different parts of the $O-C$ data, and a 4th order harmonic fit to the entire data set. The indirect period determinations were compared with directly measured period values derived for the different subsets of the data (B and D panels in Figure~\ref{fig45}). The pulsation periods were determined from the Fourier analysis of the light curve data by using the program package Mufran (Koll\'ath, 1990), whereas the periods of the modulation from the variation of the maximum brightness values. The combined, homogenized data set was divided into 7 subsets taking into account the data distribution and the pulsation period changes indicated by the maximum brightness $O-C$ shown in the top panel of Figure~\ref{fig45}. For the intervals JD 2\,436\,229 $-$ 2\,437\,463 and JD 2\,437\,652 $-$ 2\,442\,422 only the photoelectric $B$ data were analyzed. The directly determined pulsation and Blazhko periods and their errors for the 7 data subsets are listed in Table~\ref{tab2}. The good agreement between the directly measured and calculated period values confirms the correctness of our method and strengthens the validity of the period values obtained (see Figure~\ref{fig45}).
\begin{table}
\begin{tabular}{ccll}
\hline
ID & JD & \multicolumn{1}{c}{$P_0$} & \multicolumn{1}{c}{$P_{\rm{Bl}}$} \\
 & [JD $-$ 2400000] & \multicolumn{1}{c}{[day]} & \multicolumn{1}{c}{[day]} \\
\hline
1,2,3,7,8 & $17852-20985$ & 0.4680627(13) & 90.33(7) \\
4,9,10 & $23307-25441$ & 0.4670633(14) & 90.1(1) \\
10,11 & $25066-28691$ & 0.4680662(10) & 89.91(4) \\
6,12,13 & $31911-36020$ & 0.4680620(4) & 90.45(3) \\
15,17 & $36229-37463$ & 0.4680625(15) & 90.6(2) \\
15,17 & $37652-42422$ & 0.4680618(5) & 90.29(2) \\
18,19 & $47871-51633$ & 0.4680678(6) & 89.99(5) \\
\hline
\end{tabular}
\caption{Pulsation and Blazhko periods derived from the combined, homogenized subsets of the photometric data. Pulsation and modulation periods were determined from the light curves, and from the magnitudes of maxima, respectively.}
\label{tab2}
\end{table}

\section{Conclusion}
Using data of a hundred-year time base we confirm that the pulsation and modulation period changes of RV UMa are more or less the opposite of each other as it was already found in previous investigations by Bal\'azs \& Detre (1957) and Kany\'o (1976). According to the latest observations (Hipparcos, NSVS) the anti-correlated change of the periods still holds.

There are only three Blazhko variables whose long term changes in their pulsation and modulation periods could be followed. Both positive (XZ Dra: Jurcsik, Benk\H{o} \& Szeidl 2002) and negative (XZ Cyg:  LaCluzy\'e et al. 2004; \hbox{RW Dra:} Bal\'azs-Detre \& Detre 1962) correlation between the modulation and pulsation periods have been already detected in Blazhko variables. Any valid explanation of the Blazhko phenomenon should also explain this diversity.

%\newpage%%%%%%%%%%%%%%%%%%%%%%%%%%%%%%%%%%%%%%%%%%%%%%%%%%%%%%

\acknowledgements I would like to thank Dr. Johanna Jurcsik and Dr. B. Szeidl for fruitful discussions and their valuable comments. This research has made use of the SIMBAD database, operated at CDS-Strasbourg, France and the GEOS RR Lyrae database, operated at Observatoire de la C\^ote d'Azur, Nice University, France. The financial support of OTKA grant T-043504 is acknowledged.

%\appendix

%\section{This is the title of the first appendix}
%Larger tables, collections of images, spectra or similar kind of data shall be 
%presented in the appendix section rather than in the main body of the 
%text. Several appendices can be separated by the \verb+\section{+{\it title
%of appendix}\verb+}+ command. They are enclosed in the 
%\verb+appendix+ environment.

\end{document}